\documentclass[%
 reprint,
superscriptaddress,
nofootinbib,
 amsmath,amssymb,
 aps,
]{revtex4-2}

\usepackage{graphicx}
\usepackage{dcolumn}
\usepackage{bm}

\begin{document}

\newcommand{\RR}{\mathbb{R}}
\newcommand{\NN}{\mathbb{N}}
\newcommand{\CC}{\mathbb{C}}
\newcommand{\ZZ}{\mathbb{Z}}
\newcommand{\FF}{\mathcal{F}}
\newcommand{\Ncal}{\mathcal{N}}
\newcommand{\Hom}{\mathrm{Hom}}
\newcommand{\ppsi}{\left|\psi\right>}

\preprint{APS/123-QED}

\title{Anyons in a highly-entangled toric $xy$ model}

\author{Milo Moses}
\email{milo@beit.tech}
\affiliation{BEIT, Mogilska 43 31-545 Kraków, Poland}
\affiliation{Department of Mathematics, University of California, Santa Barbara, California, 93106, USA}
\author{Konrad Deka}
\email{konrad@beit.tech}
\affiliation{BEIT, Mogilska 43 31-545 Kraków, Poland}
\affiliation{Faculty of Mathematics and Computer Science, Jagiellonian University in Krakow, ul. Łojasiewicza 6, 30-348 Kraków, Poland}

\date{\today}

\begin{abstract}
While ostensibly coined in 1989 by Xiao-Gang Wen, the term ``topological order" has been in use since 1972 to describe the behavior of the classical $xy$ model. It has been noted that the $xy$ model does not have Wen's topological order since it is also subject a non-topological $U(1)$ gauge action. We show in a sense this is the only obstruction. That is, if gauge invariance is enforced energetically then the $xy$ model becomes purely topologically ordered. In fact, we show that the quantum $xy$ topological order is an infinite lattice limit of Kitaev's quantum double model applied to the group $G=\ZZ$.
\end{abstract}


\maketitle


\section{Introduction}

Topological order was first described in 1972 by Kosterlitz and Thouless \cite{kosterlitz2018ordering}, motivated by experimental observations \cite{chester1972quartz, herb1972mass}. Kosterlitz–Thouless' analysis describes states of the two dimensional $xy$ model where every point on a surface is assigned a unit tangent vector, representing its magnetic spin in the circle group $U(1)$. The regime is dominated by local ferromagnetic spin-spin interactions, with nearby particles wanting to have similar spins. That is, the spin texture desires to form a continuous vector field.

As temperature rises the vector field fails to be continuous at increasingly many points, called \textit{vortices}. The winding number around these points can either be $+1$ (vortex) or $-1$ (antivortex). Vortices and antivortices come in pairs (Fig. 1). After a certain critical temperature is reached the Kosterlitz–Thouless phase transition causes pairs to separate, giving way to a phase of matter in which vortices move freely.

\begin{figure}[b]
\includegraphics[scale=0.22]{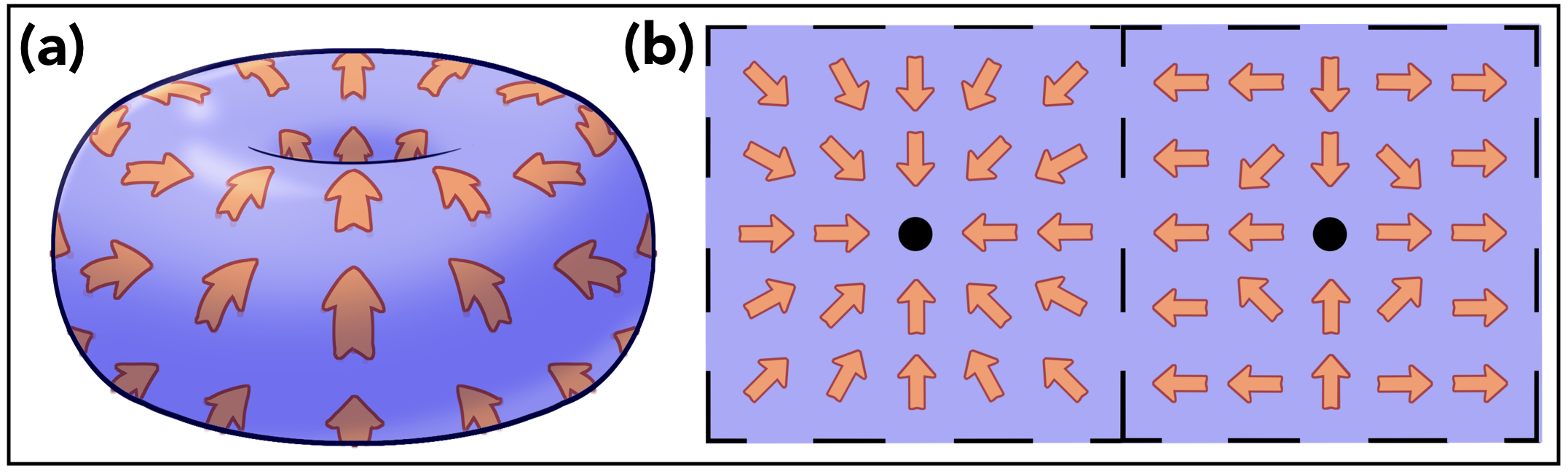}
\caption{The $xy$ model. (a) An example of a vector field on the torus, whose arrows represent directions of magnetic spins. (b) A vortex next to an antivortex. Fusing these vorticies results in an everywhere-continuous spin texture.}
\end{figure}

The difference between the classical $xy$ situation and the modern quantum situation is subtle. The term ``topological order" - ostensibly coined in 1989 by Wen \cite{wen1990topological} to describe highly entangled systems - was already in use by Kosterlitz–Thouless to describe unentangled phenomena. The difference is that $xy$ systems also have non-topological order. States are quantified not only by long-range topological properties but also by a continuous local parameter. That is, spins can always be continuously locally rotating because there is no energy cost to doing so. In fact, such fluctuations increase entropy and hence are thermodynamically favored by the Mermin–Wagner theorem \cite{mermin1966absence}.

It is an observation of Kitaev that if the $xy$ model were appropriately quantized the situation would be different. In particular, he suggests that vortices would become anyons in an $xy$ model where ``quantum fluctuations are so strong that the local order parameter is completely washed out and only the topology remains" \cite{kitaev2006anyons}. This quantization can be performed by enforcing gauge invariance energetically. That is, for every gauge transformation we add a term to the Hamiltonian which has energy $0$ if the state is gauge invariant and energy $+1$ otherwise.

In this note, we examine the behavior of the quantized $xy$-model on a torus. We show that the ground space has a basis vector corresponding to each homotopy class of non-vanishing vector fields. That is, basis vectors are superpositions over equivalence classes of vector fields which can be continuously deformed from one to another. Homotopy classes of non-vanishing vector fields on the torus $T$ correspond canonically to elements of the cohomology group $H^1(T;\ZZ)\cong \ZZ\times \ZZ$. Thus, this system has an infinite dimensional but discrete topological ground state degeneracy.

Our main result is to show that this system is an infinite lattice limit of Kitaev's quantum double model applied to the group $G=\ZZ$ \cite{kitaev2003fault}. This is done by equipping the torus with a lattice and modding out by homotopy equivalence within each face. The resulting system has an orthonormal basis consisting of assignments of integers to every edge on the lattice. These integers correspond to winding numbers. We show that local deformations ($U(1)$ gauge action) correspond to the ``$A_v$" vertex stabilizers in Kitaev's model, and the continuity condition on the vector field corresponds to the ``$B_p$" face stabilizers.

While the results of this paper are only stated for the case of the $xy$ model, the result illustrated holds in the much more general context of ordered media \cite{mermin1979topological}. Suppose that $R$ is some smooth manifold representing order parameter space. The defects in ordered media behave behave to a large degree like anyons. There are a few key differences. One serious difference is the possibility of $\pi_2(R)\neq 0$. This would means that vaccum quasiparticle has non-trivial morphisms to itself, which is typically disallowed. It is thus natural to restrict ourselves to the case $\pi_2(R)=0$. At this point, the only difference between defects and anyons is that there is also a non-topological order paramater present in the system. This can be circumvented using the same construction as above, enforcing gauge invariance energetically. The resulting topological order will be equivalent to the Kitaev quantum double model with gauge group $G=\pi_1(R)$, though we do not prove it here. Applying this construction to the Eilenberg-Maclane order paramter space $K(G,1)$ in higher dimensions recovers higher dimensional Dijkgraaf-Witten theory \cite{freed2009topological}. Note that in these cases we work with the trivial bundle instead of the tangent bundle so our construction can be applied to real-space manifolds other than the torus.

We additionally observe that Kitaev's original model is only defined for finite groups, since in the infinite case mathematical issues arise. Namely, the $A_v$ stabilizers don't have literal eigenvectors so we must resort to distributions to make a coherent theory. A more natural setting in which to describe these sorts of non-finite topological in category-theory, as has been done in Ref. \cite{geer2022pseudo}, where our theory arises implicitly as a special case of pseuo-Hermitian Levin-Wen models.

Like with finite groups, $G=\ZZ$ anyons come in two classes. The first class corresponds to failures of $B_p$ stabilizers, for which there is one type for every integer in $\ZZ$. These are the vortices. The presence of a quasiparticle of type $n\in \ZZ$ that a face contains $|n|$ vortices/antivorticies. The second class of anyons has one type for every phase in $U(1)$.

Vortices exhibit Bosonic exchange statistics with one-another, but anyonic statistics when braided with $U(1)$-type quasiparticles. When quasiparticles of type $n\in\ZZ$ and $\lambda\in U(1)$ are braided, the resulting phase factor is $\lambda^n$. Vortices also exhibit non-trivial actions on the ground space when moved around homotopically non-trivial loops.

Lastly, we offer a comment on measurement. Observing the exact direction of each spin is impossible. However what can be determined are the topologically-protected winding numbers around axes of the torus, by sending around electrons. When an electron moves adiabatically it acquires a Berry phase, equal to $-1$ if the total winding number along its trajectory is odd and $+1$ if it is even \cite{cohen2019geometric}. Performing interferometry, such phases allow us to measure the winding number modulo $2$ - it is not clear whether there is a measurement technique which distinguishes between all winding numbers.

\section{Lattices}

We now demonstrate the procedure of going from a vector field model a to lattice based model. Given a non-vanishing vector field we assign discrete data on a lattice as follows:

\begin{enumerate}
\item Equip the torus with a lattice structure.
\item Locally twist the field so that the vectors at the vertices all point in the same direction.\footnote{This step can be abstractly viewed as choosing a trivialization of the tangent bundle, which is algebraically necessary for identifying vector fields with cocycles.}

\item On every edge write the total winding number.
\end{enumerate}

The winding numbers on the boundries of the faces uniquely specify homotopy classes of fillings, since the faces are contractible and $\pi_2(U(1))=0$. This procedure is thus in a sense modding out by local deformations within faces of the lattice. As the lattice gets finer this data specifies vector fields up to increasingly local deformations. We give an example in Fig.2.

\begin{figure*}
\includegraphics[scale=0.4]{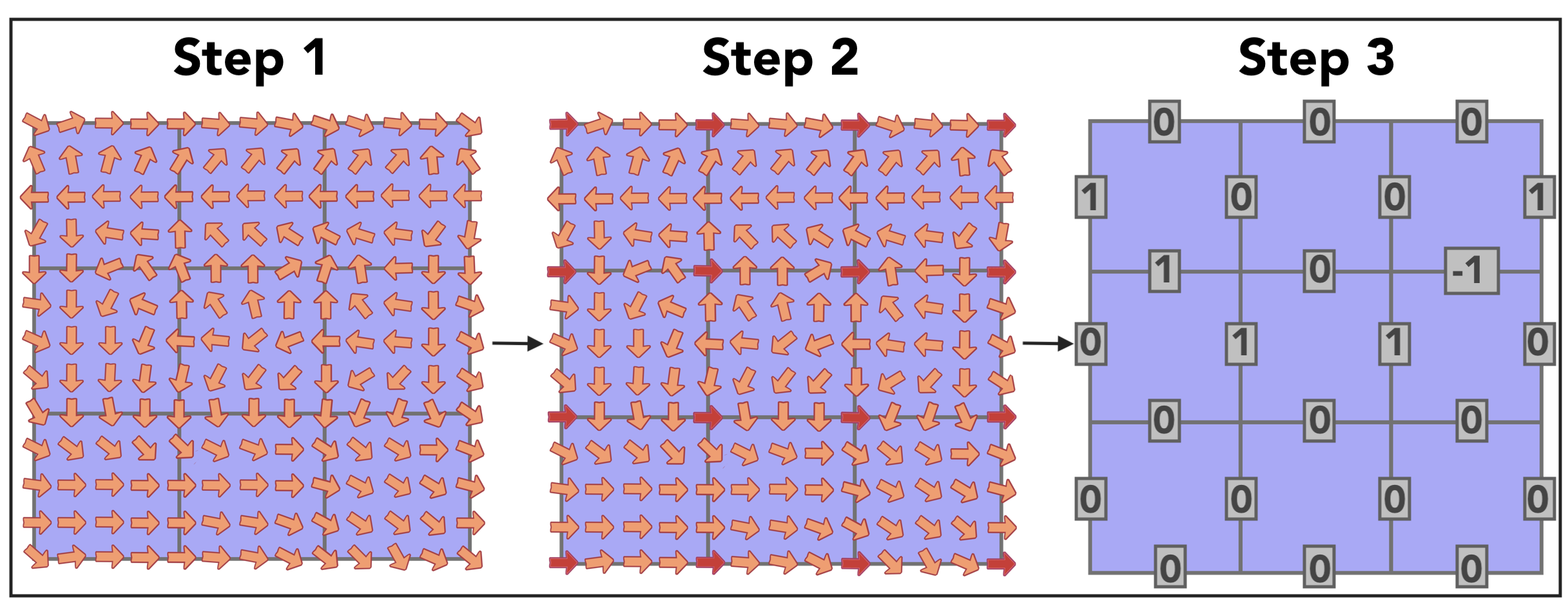}
\caption{Going from a vector field to integers on a lattice. In Step 1 we draw a lattice on the torus, which we have identified here by its gluing diagram. In Step 2, we make local twists so that all vectors at vertices point in the same direction. In Step 3, we record the total winding numbers along each edge.}
\end{figure*}

Given integers assigned to each lattice edge there won't always exist a vector field whose winding numbers along edges gives the integers specified. However, there is a simple condition to check: a face with integers assigned to its edges is fillable by a vector field if and only if the sums of the integers along the boundary is $0$ \footnote{Of course, one must be careful about sign conventions for clockwise/counterclockwise rotations and add signs accordingly.}. We thus have arrived at a discrete lattice-based way of thinking about vector fields - they are assignments of integers to lattice edges such that the signed sum around each face is 0.

This leaves open the question of deformations. What effect can continuous local changes have on edge data? If the deformation happens at a face or edge, no winding numbers will change. However, if it happens at a vertex this can change winding numbers as seen in Fig.3. Specifically, rotations have the effect of adding $\pm n$ to the surrounding edges depending on orientation. Local rotations are a $U(1)$ gauge action.

\begin{figure}[b]
\includegraphics[scale=0.28]{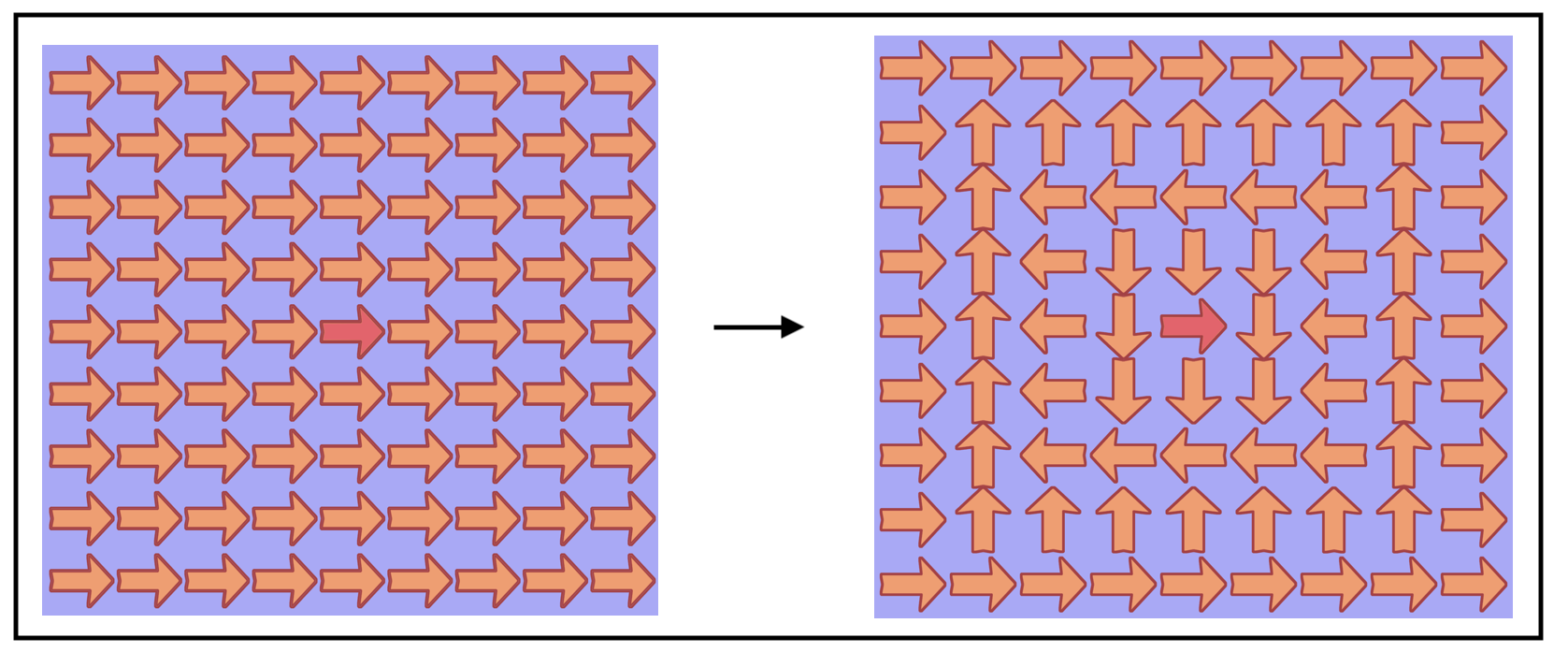}
\caption{A local rotation of a vector field at a vertex. This changes the winding numbers of all the edges touching it.}
\end{figure}

The quantzied version of this system allows superpositions of vector fields. That is, instead of putting integers on edges we put an element of the Hilbert space $L^2(\ZZ)$ with infinite discrete orthonormal basis $\{\left| n\right>,\,\, n\in \ZZ\}$. Ground states are those whose support consists of assignments whose signed sum is $0$ along every face, and which are invariant under the $U(1)$ gauge action at every vertex.

\section{Algebraic Description}

We introduce some notation. The quantum system we are working in is

$$\Ncal=\bigotimes_{\text{edges}}L^2(\ZZ).$$

For all $n\in\ZZ$, $\lambda\in U(1)$ we define ``translation" and ``rotation" maps

\begin{align*}
T_n:L^2(\ZZ)&\xrightarrow{}L^2(\ZZ),\,\,\,\,\,\,\,\,\,\,\, R_\lambda:L^2(\ZZ)\xrightarrow{}L^2(\ZZ).\\
\left| m\right>&\mapsto \left| m+ n\right>\,\,\,\,\,\,\,\,\,\,\,\,\,\,\, \left| n\right>\mapsto \lambda^n\left| n\right>
\end{align*}

For all vertices $v$ and faces $p$ we define operators

$$A_{n,v}=\bigotimes_{\text{edges touching }v}T_{\pm n},\,\,\,\,\,\, B_{\lambda,p}=\bigotimes_{\text{edges bounding }p}R_{\lambda^{\pm 1}}$$

where the signs depend on relative edge orientations. We let $A_v$ be the projector into the simultaneous $+1$ eigensapce of $A_{n,v}$ and we let $B_{p}$ be the projector into the simultaneous $+1$ eigenspace of the $B_{\lambda,p}$ operators. The Hamiltonian for our system is

$$H=\sum_{\text{verticies  }v}(1-A_v)+\sum_{\text{faces }p}(1-B_p).$$

Since $[A_v,B_p]=0$, the ground state of the system corresponds to those in the simultanous $+1$ eigensapce of all $A_{n,v}$ and $B_{\lambda,p}$. Being in the $+1$ eigenspace of the $A_{n,v}$ is gauge invariance, and being in the $+1$ eigenspace of the $B_{\lambda,p}$ is continuity.

It is clear by our construction that ground states will correspond to homotopy classes of vector fields on the torus $T$. Explicitly, every lattice cocycle in $ C^1(T;\ZZ)$ gives an assignment integers to edges, and the cocycle condition guarantees being in the $+1$ eigenspace of the $B_{\lambda,p}$ operators. Taking an equal superposition over cohomologous cocycles gives a gauge invariant state. Hence there is a ground state for every cohomology class in $H^1(T;\ZZ)$, so the ground space is canonically isomorphic to $\CC[H^1(T;\ZZ)]$. These infinite linear combinations are not literal elements of the Hilbert space - they're added to the theory as distributions.

Quasiparticles are failures of stabilizers. That is, faces or vertices at which stabilizers don't act by $1$. Different quasiparticle types correspond to different stabilizer eigenvalues\footnote{At least, this is true in the abelian case. Quasiparticles in non-abelian gauge theories can correspond to higher dimensional irreducible representations.}. At faces, quasiparticles are familiar: they're vortices. Namely, if a state $\ppsi\in \Ncal$ is a common eigenvector of the stabilizers then for every face $p$ there exists a unique $n\in\ZZ$ such that

$$B_{\lambda,p}\ppsi=\lambda^n\ppsi,\,\, \forall \lambda\in U(1).$$

We say in this case that there is an $n$-type quasiparticle at $p$. This value $n$ is the total winding number around $p$, and hence indicates the presence of $|n|$ vortices/antivorticies.

\section{Vertex quasiparticles}

Vertex quasiparticles correspond to eigenvalues of the maps $A_{n,v}$. As stated before the $A_{n,v}$ operators don't have eigenvalues in the usual sense. However, we do have a Fourier series map

\begin{align*}
\FF: L^2(\ZZ)&\xrightarrow{\sim}L^2(U(1)).\\
\sum_{m\in\ZZ}c_m \left |m\right> &\mapsto \left(\lambda\mapsto \sum_{m\in\ZZ}c_m \lambda^m\right)
\end{align*}

The action of $T_n$ on $L^2(U(1))$ is multiplication by $\lambda^n$. In a sense this is the position operator, whose eigenvectors are Dirac deltas. Passing back through the Fourier transform, we thus find generalized eigenstates

$$\sum_{m\in\ZZ}\lambda^{-m}\left| m\right>,\,\, \lambda\in U(1).$$

The eigenvalue of the $\lambda$-state under $T_n$ is $\lambda^n$. The eigenvalues of $A_{n,v}$ thus all lie in $U(1)$. If a state $\ppsi\in \Ncal$ is a common eigenvector of the stabilizers, then there exists a unique $\lambda\in U(1)$ such that

$$A_{n,v}\ppsi=\lambda^n\ppsi,\,\, \forall n\in\ZZ.$$

In this case, we say there is a $\lambda$-type quasiparticle living at $v$. These quasiparticles are phase failures, spread out over the entangled deformations of vector fields.

\section{Braiding}

The key commutator relationship in the braiding theory is

$$R_\lambda T_n=\lambda^nT_nR_\lambda.$$

It immediately implies that $R_{\lambda} A_{n,v} =\lambda^{\pm n}A_{n,v}R_{\lambda}$ if $R_{\lambda}$ is applied to an edge touching $v$, and $R_{\lambda}A_{n,v}=A_{n,v}R_{\lambda}$ otherwise. In other words, applying $R_{\lambda^{\pm 1}}$ moves $\lambda$-type quasiparticles along the lattice. Similarly, applying $T_{n}$ moves $n$-type quasipartricles along the dual lattice, coming from the relation $T_n B_{\lambda,p}=\lambda^{\pm n} B_{\lambda,p}T_n$ whenever $T_n$ is applied to an edge bounding the face $p$.

If a $\lambda$-type quasiparticle is braided around a $n$-type quasiparticle the resulting phase factor is $\lambda^n$. The only other topologically non-trivial procedure that can be performed is moving quasiparticles around non-trivial loops on the torus, which has a predicatable action on the ground states whose description we omit.
$\newline$

\begin{acknowledgments}
We thank all of our amazing coworkers at BEIT for making this work possible, especially Emil Żak for his invaluable physical insights, and Jacek Horecki for his algebraic-topological expertise. M.M. would like to thank Nathan Geer for bringing to his attention reference \cite{geer2022pseudo}, and Jinwen Zhao for making the figures in this note. This project was co-funded by the European Union. Views and opinions expressed are however those of the authors only and do not necessarily reflect those of the European Union or European Innovation Councien and SMEs Executive Agency (EISMEA). Neither the European Union nor the granting authority can be help responsible for them
\end{acknowledgments}

\bibliography{apssamp}

\end{document}